# Diffraction modelling of a 2023 March 5 stellar occultation by subkilometer-sized asteroid (98943) 2001 CC21

Ko Arimatsu ,[1,*] Fumi Yoshida ,[2,3] Tsutomu Hayamizu,[4] Miyoshi Ida,[5] George L. Hashimoto ,[6] Takashi Abe,[7] Hiroshi Akitaya ,[8,3,9] Akari Aratani,[10] Hidekazu Fukuda,[11] Yasuhide Fujita,[12] Takao Fujiwara,[6] Toshihiro Horikawa,[7] Tamio Iihoshi,[13] Kazuyoshi Imamura,[14] Ryo Imazawa ,[15] Hisashi Kasebe,[5] Ryosuke Kawasaki,[16] Hiroshi Kishimoto,[5] Kazuhisa Mishima,[17] Machiko Miyachi,[18] Masanori Mizutani,[19] Maya Nakajima,[6] Hiroyoshi Nakatani,[13] Kazuhiko Okamura,[7] Misaki Okanobu,[6] Masataka Okuda,[20] Yuji Suzuki,[21] Naoto Tatsumi,[22] Masafumi Uno,[7] Hidehito Yamamura,[23] Mikoto Yasue,[24,10] Hideki Yoshihara,[5] Masatoshi Hirabayashi ,[25] and Makoto Yoshikawa [26]

[1] The Hakubi Center/Astronomical Observatory, Graduate School of Science, Kyoto University Kitashirakawa-oiwake-cho, Sakyo-ku, Kyoto, Kyoto 606-8502, Japan
[2] University of Occupational and Environmental Health, Japan, 1-1 Iseigaoka, Yahatanishi, Kitakyusyu, Fukuoka 807-8555, Japan
[3] Planetary Exploration Research Center, Chiba Institute of Technology, 2-17-1 Tsudanuma, Narashino, Chiba 275-0016, Japan
[4] Saga Hoshizora Astronomy Center, 328 Nishiyoka-cho, Oaza Takataro, Saga, Saga 840-0036, Japan
[5] Japan Occultation Information Network (JOIN), Japan
[6] Department of Earth Sciences, Okayama University, 3-1-1 Tsushimanaka, Kita, Okayama, Okayama 700-8530, Japan
[7] Museum of Astronomical Telescopes, 30-1 Sukemitsu-Higashi, Tawa, Sanuki, Kagawa 769-2306, Japan
[8] Astronomy Research Center, Chiba Institute of Technology, 2-17-1 Tsudanuma, Narashino, Chiba 275-0016, Japan
[9] Hiroshima Astrophysical Science Center, Hiroshima University, 1-3-1 Kagamiyama, Higashi-Hiroshima, Hiroshima 739-8526, Japan
[10] GROUND Geoscience club, Hokkaido University, Kita-10 Nishi-8, Kita-ku, Sapporo, Hokkaido 060-0810, Japan
[11] Kakogawa Outdoor Education Center, 715-5 Amagahara, Higashikanki-cho, Kakogawa, Hyogo 675-0058, Japan
[12] Kuma Kogen Astronomical Observatory, 488 Shimohatanokawa Otsu, Kumakogen, Kamiukena-gun, Ehime, 791-1212, Japan
[13] Astronomical Club of Fukuyama, Fukuyama, Hiroshima, Japan
[14] Anan Science Center, 8-1 Nagakawa Kamifukui Minami-Kawabuchi, Anan, Tokushima 779-1243, Japan
[15] Department of Physics, Graduate School of Advanced Science and Engineering, Hiroshima University, 1-3-1 Kagamiyama, Higashi-Hiroshima, Hiroshima 739-8526, Japan
[16] Kougominami, Nishi-ku, Hiroshima, Hiroshima 733-0823, Japan
[17] Life Park Kurashiki Science Center, 940 Koshinden, Fukudacho, Kurashiki, Okayama 712-8046, Japan
[18] Japanese Association of Accounting and Astronomy, 1-1-1 Nishimachi, Fukuyama, Hiroshima 720-0067, Japan
[19] Hattoji Star Watching Group, 1393 Kagami, Yoshinaga-cho, Bizen, Okayama 709-0301, Japan
[20] Kaichi, Yao, Osaka 581-0881, Japan
[21] Ehime Prefecture Science Museum, 2133-2 Ojoin, Niihama, Ehime 792-0060, Japan
[22] Ryuten Astronomical Observatory, 2978-3 Nakaseijitsu, Akaiwa, Okayama 701-2437, Japan
[23] NPO KWASAN ASTRO NETWORK, c/o Kwasan Observatory, 17 Kitakazan-ohmine-cho, Yamashina-ku, Kyoto, Kyoto 607-8471 Japan
[24] Hokkaido University Astronomy Club, Hokkaido University, Kita-10 Nishi-8, Kita-ku, Sapporo, Hokkaido 060-0810, Japan
[25] Daniel Guggenheim School of Aerospace Engineering, Georgia Institute of Technology, North Avenue, Atlanta, GA 30332, USA
[26] Institute of Space and Astronautical Science (ISAS), Japan Aerospace Exploration Agency (JAXA), 3-1-1 Yoshinodai, Chuo-ku, Sagamihara, Kanagawa 252-5210, Japan
*E-mail: arimatsukou@gmail.com

## Abstract
We present an analysis of a stellar occultation event caused by a near-Earth asteroid (98943) 2001 CC21, an upcoming flyby target in the Hayabusa2 extended mission, on 2023 March 5. To determine the asteroid's shape from diffraction-affected light curves accurately, we developed a novel data-reduction technique named the Diffracted Occultation's United Simulator for Highly Informative Transient Explorations (DOUSHITE). Using DOUSHITE-generated synthetic models, we derived constraints on (98943) 2001 CC21's shadow shape from the single-chord occultation data. Our results suggest a significant elongation of the shadow with an axis ratio of $b/a = 0.37 \pm 0.09$. This shape could be crucial for planning Hayabusa2's high-speed flyby to optimise the limited imaging opportunities.

# 1 Introduction

Stellar occultations by small solar system bodies have been used for the past 60 years to determine their orbits, sizes, and shapes (e.g., Sinvhal et al. 1962). The shadow length for each occultation chord is generally determined by estimating the immersion and emersion times of the event. These times are estimated by fitting a synthetic light curve that either excludes diffraction effects or includes the Fresnel diffraction fringe of an infinite, straight edge of a geometrical shadow convolved by a stellar size. These classical methods can provide rough information about the geometrical shape of an occulting body at distance $D$ from the observer if the body size is significantly larger than the Fresnel scale $f$, which is given by

$$f = \sqrt{\frac{D\lambda}{2}} \sim 0.19 \sqrt{\frac{D}{1\,\mathrm{au}}}\ \mathrm{km}, \qquad (1)$$

where $\lambda$ is the wavelength of the observations and is assumed to be $\lambda = 500$ nm.

Recent developments in stellar astrometry, such as the Gaia catalog (e.g., Gaia Collaboration 2021, 2023), have allowed for unprecedented accuracy in predicting occultation events by kilometer- to subkilometer-sized bodies (e.g., Yoshida et al. 2023). In addition, several observation programmes for small, very distant, and unknown objects will be performed in the near future (Arimatsu et al. 2017, 2019b). However, the sizes of these occulting bodies are close to the Fresnel scale, and the light curves suffer severe diffraction effects. To constrain the shapes of the occultation shadows with limited observational resources, an alternative data reduction technique is necessary.

On 2023 March 5, a stellar occultation event by the asteroid (98943) 2001 CC21 occurred in western Japan. (98943) 2001 CC21 is a near-Earth asteroid (NEA) belonging to the Apollo group and has been selected as a flyby target of JAXA's Hayabusa2 extended mission (Hayabusa2#, Hirabayashi et al. 2021). In 2026 July, the spacecraft will approach the asteroid at a distance of less than 100 km, with a relative velocity of $\sim 5$ km s$^{-1}$. Due to the limited time available during the flyby, it will be challenging to make a detailed observation plan of the asteroid based on in situ measurements. Accurate shape estimates before the flyby help to plan the proximity observations on the Hayabusa2# extended mission.[1]

This paper proposes a new data-reduction technique for stellar occultations by small or distant solar system objects. The technique considers the diffraction effect produced by occulting shadows of arbitrary shape in the observed light curves. Additionally, a Bayesian approach is adopted to derive posterior probability distributions for the size, shape, and orientation parameters from the limited datasets. The methodology for modeling diffraction effects is presented in section 2. The observations and data reductions of the 2023 March 5 stellar occultation event is described in section 3. The results of the data reductions and the model comparisons are given in section 4. The obtained shape parameters of (98943) 2001 CC21 are discussed in section 5. Finally, section 6 provides a summary of the results and discussions.

# 2 Diffraction modelling method

To provide accurate model light curves of stellar occultation events, we have developed a diffraction modelling procedure called Diffracted Occultation's United Simulator for Highly Informative Transient Explorations (DOUSHITE). DOUSHITE was initially designed for stellar occultation events caused by unidentified trans-Neptunian objects (TNOs) observed with high-cadence multiple observation systems such as Organized Autotelescopes for Serendipitous Event Survey (OASES; Arimatsu et al. 2017, 2019b), and also those caused by known larger TNOs that possibly have atmospheres (e.g., Arimatsu et al. 2019a, 2020). In the DOUSHITE procedure, synthetic light curves of occultation events induced by occulting shadows of an arbitrary shape are generated, taking into account the diffraction effects with the techniques proposed by Roques et al. (1987).

We consider the diffraction by a section of the object $S$ at distance $D$ with an arbitrary two-dimensional shape in the $X$–$Y$ plane consisting of small rectangular screen elements (figure 1). The minimum and maximum coordinates of a screen element $k$ are expressed as $(x_{k1}, y_{k1})$ and $(x_{k2}, y_{k2})$, respectively. When both the dimensions of $S$ are much smaller than $D$, i.e., $(x_{k1} - x)^2 + (y_{k1} - y)^2 \ll D^2$ and $(x_{k2} - x)^2 + (y_{k2} - y)^2 \ll D^2$ for all $k$, the complex diffraction amplitude $a(x, y; \lambda)$, at the observer's plane $(x, y)$, at wavelength $\lambda$, is given from the Fresnel–Kirchhoff diffraction theory by

$$a(x, y; \lambda) = 1 - \frac{1}{i\lambda D} \int_{x_{k1} - x}^{x_{k2} - x} \exp\left(\frac{i\pi}{\lambda D} X^2\right) \mathrm{d}X$$

$$\times \int_{y_{k1} - y}^{y_{k2} - y} \exp\left(\frac{i\pi}{\lambda D} Y^2\right) \mathrm{d}Y. \qquad (2)$$

If we adopt the Fresnel scale $f = \sqrt{(D\lambda/2)}$ (equation 1) as a normalization factor, we obtain

$$a(x, y; \lambda) = 1 - \frac{1}{2i}\left[F(u_{k2}) - F(u_{k1})\right]$$

$$\times \left[F(v_{k2}) - F(v_{k1})\right], \qquad (3)$$

where $F(u)$ is the complex Fresnel integral given by

$$F(u) = \int_0^u \exp\left(\frac{i\pi}{2} t^2\right) \mathrm{d}t, \qquad (4)$$

and $u_{k1}$, $u_{k2}$, $v_{k1}$, and $v_{k2}$ are normalized variables defined by $u_{k1} = (x_{k1} - x)/f$, $u_{k2} = (x_{k2} - x)/f$, $v_{k1} = (y_{k1} - y)/f$, and $v_{k2} = (y_{k2} - y)/f$, respectively. The complex amplitude $A_S(x, y; \lambda)$, diffracted by the entire shadow $S$, at wavelength $\lambda$ can be expressed as the sum of those of the individual elements and is obtained as

$$A_S(x, y; \lambda) = 1 - \sum_k \frac{1}{2i}\left[F(u_{k2}) - F(u_{k1})\right]$$

$$\times \left[F(v_{k2}) - F(v_{k1})\right]. \qquad (5)$$

The monochromatic diffraction intensity at $(x, y)$, $J(x, y; \lambda)$ is given with the initial intensity $J_0$ by

$$J(x, y; \lambda) = J_0\,|A_S(x, y; \lambda)|^2. \qquad (6)$$

In the following, we consider light curves that are scaled to the stellar flux observed before and after the occultation event. Therefore, we set $J_0 = 1$.

In general, occultation events are observed with instruments covering a range of wavelengths. The diffraction pattern is

---

[1] ⟨https://www.hayabusa2.jaxa.jp/en/⟩.

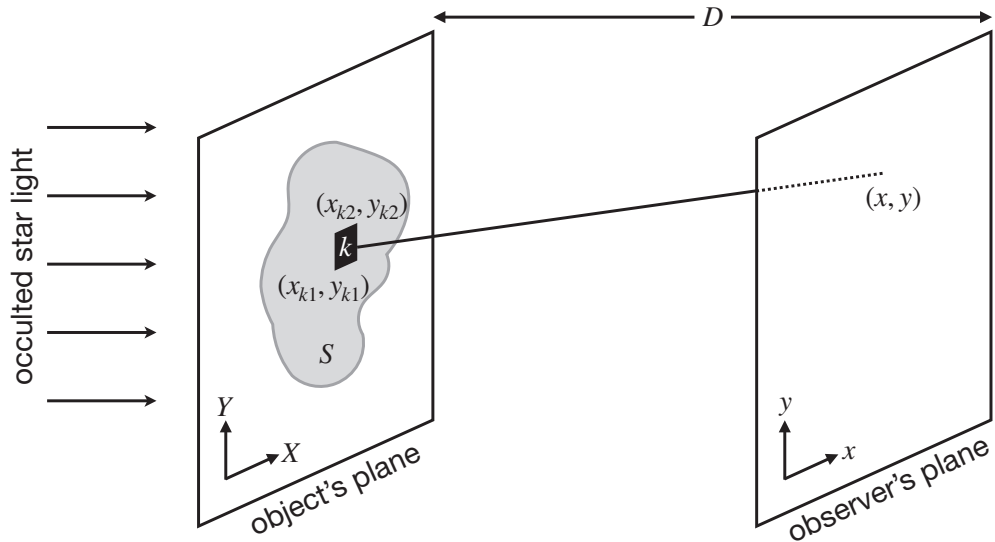

**Fig. 1.** Notation of the geometry used in the Fresnel diffraction model. The rectangular screen element $k$ constitutes the object's shadow $S$ in the $X$–$Y$ plane. The $x$–$y$ plane represents the observer's plane. The diffraction amplitude at $(x, y)$ generated by $k$ is calculated using the Fresnel–Kirchhoff diffraction theory. See section 2.

thus obtained by averaging $J(x, y; \lambda)$ over wavelengths as

$$J(x, y) = \frac{\int R(\lambda) J(x, y; \lambda) \, d\lambda}{\int R(\lambda) \, d\lambda}. \quad (7)$$

$R(\lambda)$ is the system response at wavelength $\lambda$ for the observation and is given by

$$R(\lambda) = s(\lambda) e(\lambda), \quad (8)$$

where $s(\lambda)$ is the spectrum of the occulted star and $e(\lambda)$ is the spectral efficiency of the observing instrument.

For direct comparison with observed light curves, the synthetic light curve $L(t)$, at time $t$, is obtained by convolving $J(x, y)$ with a stellar disk of a finite angular size and is given by

$$L(t) = \frac{\iint \phi(x, y; t) J(x, y) \, dx \, dy}{\iint \phi(x, y; t) \, dx \, dy}, \quad (9)$$

where $\phi(x, y; t)$ is the convolution kernel representing the finite-sized stellar disk at $t$. This study does not take into account the limb darkening effect of the stellar disk, which could affect the resulting light curve with an amplitude significantly less than 1%.

Additionally, it is crucial to incorporate the effects of finite integration time on the synthetic light curve model. We therefore define $L_i$ for the data point $i$ whose beginning and ending integration times are $t_{i\min}$ and $t_{i\max}$, respectively. $L_i$ is expressed as

$$L_i = \int_{t_{i\min}}^{t_{i\max}} L(t) \, dt. \quad (10)$$

## 3 Observation and data reduction

The occultation of the star TYC 4082-00763-1 (Gaia DR3 catalog source ID: 472569655241986816, ICRS position at epoch J2000.0: $\alpha = 4^{\rm h} 40^{\rm m} 36{.}^{\rm s}7$, $\delta = +62°15'40''$, Gaia $G$-band magnitude = 10.1; Gaia Collaboration 2023) by (98943) 2001 CC21 was predicted to occur on 2023 March 5, according to the prediction calculated using OCCULT version 4 software[2] with JPL201 orbital elements and the Gaia EDR3 stellar catalog (Gaia Collaboration 2021). The prediction indicated that (98943) 2001 CC21's shadow would pass over the Honshu and Shikoku islands in Japan at an apparent speed of $\sim$4.3 km s$^{-1}$. The predicted path map of the occultation shadow is shown in figure 2. The geocentric distance $D$ of (98943) 2001 CC21 during the occultation is $D = 0.1313$ au. It should be noted that a quality indicator for the astrometric accuracy of the Gaia EDR3 catalog, the renormalised unit weight error (RUWE), for TYC 4082-00763-1 (RUWE = 2.10) is larger than the conventional threshold (RUWE = 1.4; Ferreira et al. 2022). Therefore, the present occultation event is not suitable for determining the astrometry of the asteroid.

A coordinated observation campaign for the 2023 March 5 event was carried out at 20 observing sites in Japan. Table 1 summarizes the location (latitude, longitude, and elevation), the observing equipment [telescope aperture, complementary metal-oxide semiconductor (CMOS) camera, exposure time, and time acquisition method], and the observation status for each observing site. The locations of the observing sites are shown in figure 2. Of the 20 sites, 18 were able to make photometric observations of the star during the predicted time ranges of occultation. Figure 3 shows the occultation chord map for the 18 successful observing sites in the projected sky plane.

Most CMOS cameras do not have the capability to record time directly. Therefore, the capture time of each frame is obtained from the timestamp information added to the image data by the capture software. However, as noted in previous occultation studies (e.g., Arimatsu et al. 2017; Yoshida et al. 2023), timestamps recorded in images can be unreliable, possibly caused by a time delay between the time the image is captured and when the image is recorded on the storage devices, even in the case where the control PC system clock is synchronized with a Global Positioning System (GPS) receiver. To solve this problem, we used the one pulse-per-second (1 PPS) emission of the light-emitting diode (LED) generated by the GPS module and projected on to the image. The detailed concept of precise time-measurement using the 1 PPS LED emission is presented in Yoshida et al. (2023). We measured the difference between the time stamp of the image and the timing of the 1 PPS emission using a light curve of the LED. To avoid rolling shutter effects of the image sensors, we measured the LED emission in the pixels located in the same row as that of the occulted star. The time stamp was corrected by subtracting the time difference. These time-correction procedures follow those implemented in the occultation data analysis software "Limovie."[3]

---

[2] ⟨http://lunar-occultations.com/iota/occult4.htm⟩.

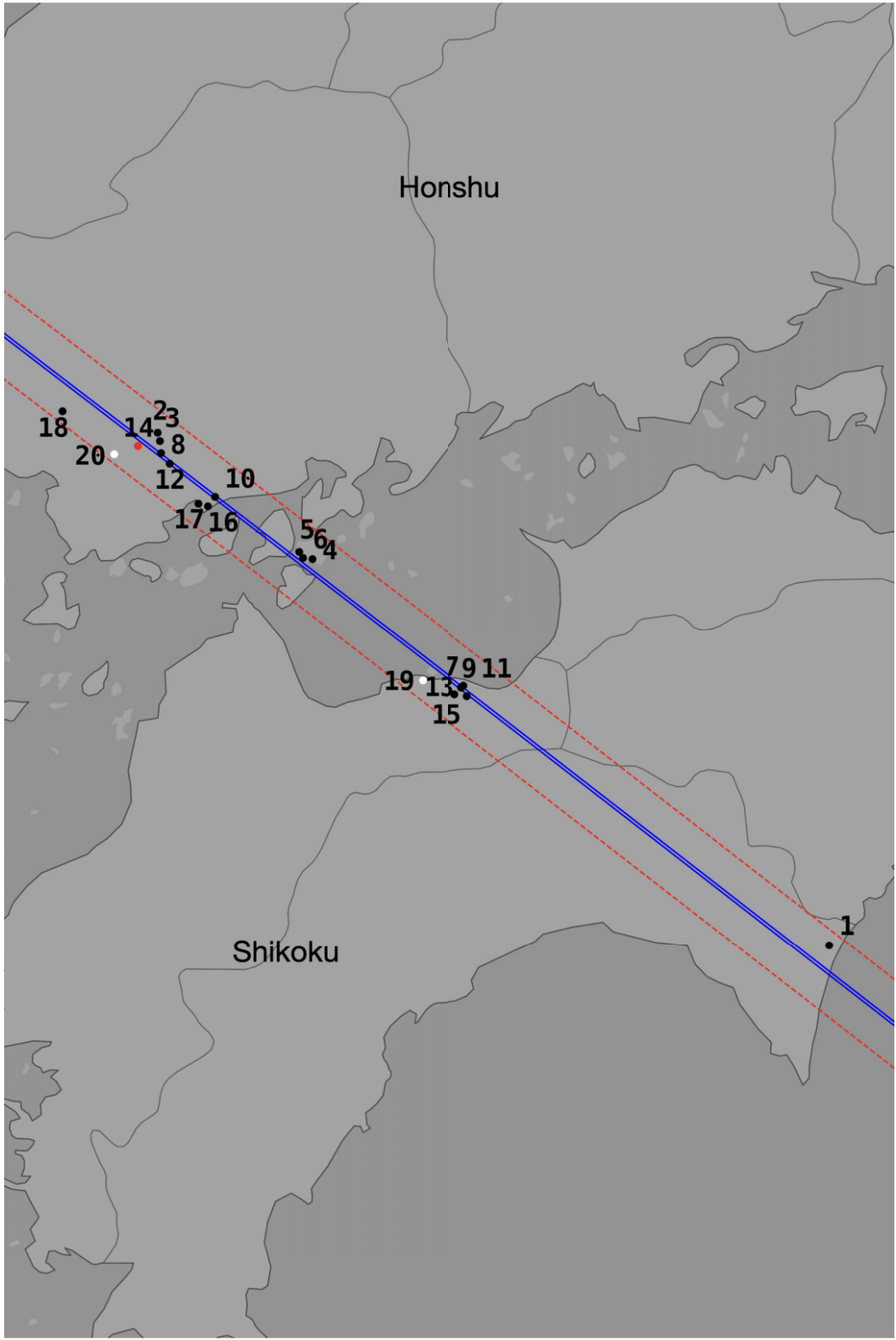

**Fig. 2.** Predicted occultation shadow path map of the 2023 March 5 (98943) 2001 CC21 event overlaid with the location of the observation sites. The blue lines indicate the predicted path of the occultation shadow, and the red dashed lines represent the 1$\sigma$ error range. The red circle indicates the location of the positive detection (Site No. 14). The black points represent the sites of negative detection, i.e., (98943) 2001 CC21 did not occult the target star. The white dots indicate the locations where observations were unsuccessful due to instrument trouble.

**Table 1.** Observation array.

| Obs. site | Observers | Longitude | Latitude | Alt. [m] | Tel. ap. [cm] | Camera | Exp. time [ms] | Time keeping | Status[*] |
|---|---|---|---|---|---|---|---|---|---|
| 1 | K. Imamura | 134°14′9″.6 | 33°29′33″.7 | 37 | 25.4 | ASI290MM | 30 | GPS | N |
| 2 | Yasue et al. | 132°44′59″.3 | 34°26′2″.7 | 221 | 20.3 | ASI290MM | 34.4 | GPS | N |
| 3 | T. Iihoshi | 132°45′19″.5 | 34°25′08″.2 | 239 | 20 | ASI533MC Pro | 20.8 | GPS | N |
| 4 | M. Okuda | 133°5′39″.7 | 34°12′12″.8 | 3 | 20 | ASI1600MC-cool | 30 | GPS | N |
| 5 | H. Kasebe | 133°3′56″.5 | 34°13′0″.4 | 3 | 20 | ASI290MM | 20.8 | GPS | N |
| 6 | H. Kishimoto | 133°4′26″.0 | 34°12′20″.6 | 3 | 21 | ASI290MM | 25.6 | GPS | N |
| 7 | T. Abe | 133°25′57″.5 | 33°58′18″.4 | 8 | 20 | ASI294MC | 25.6 | GPS | N |
| 8 | H. Fukuda | 132°45′27″.0 | 34°23′49″.9 | 218 | 20 | ILCE-7S | 33.3 | GPS | N |
| 9 | Suzuki et al. | 133°25′40″.0 | 33°58′4″.3 | 11.6 | 27.9 | ASI2600MM Pro | 32.2 | GPS | N |
| 10 | M. Mizutani | 132°52′42″.4 | 34°19′01″.9 | 3.3 | 20 | ASI178MM | 23 | GPS | N |
| 11 | K. Mishima | 133°25′40″.2 | 33°58′4″.6 | 11.6 | 18 | WAT-910HX/RC | 16 | GPS | N |
| 12 | Akitaya et al. | 132°46′37″.2 | 34°22′39″.7 | 512 | 150 | QHY174M GPS | 20 | GPS | N |
| 13 | K. Okamura | 133°26′25″.6 | 33°57′8″.7 | 110 | 16 | ASI294MC Pro | 31.2 | GPS | N |
| 14 | M. Ida | 132°42′24″.5 | 34°24′36″.1 | 217 | 20 | ASI290MM | 25.6 | GPS | P |
| 15 | T. Horikawa | 133°24′45″.8 | 33°57′20″.0 | 33.8 | 20 | SV305 | 34.4 | GPS | N |
| 16 | M. Uno | 132°51′45″.3 | 34°17′58″.9 | 3 | 20 | QHYIII178C | 30.4 | GPS | N |
| 17 | H. Yoshihara | 132°50′28″.7 | 34°18′16″.9 | 13.5 | 20 | ASI290MM | 30.4 | GPS | N |
| 18 | T. Hayamizu | 132°32′17″.5 | 34°28′19″.9 | 61 | 20 | WAT-100N | 33 | GPS | N |
| 19 | H. Yamamura | 133°20′23″.7 | 33°58′52″.3 | 3 | 20 | ASI290MM | 31.2 | GPS | F |
| 20 | N. Tatsumi | 132°39′13″.8 | 34°23′38″.0 | 269 | 20 | SV305-SJ | 30.6 | GPS | F |

[*] "P" and "N" are positive detection and negative detections of the occultation, respectively. "F" is an instrument failure or trouble.

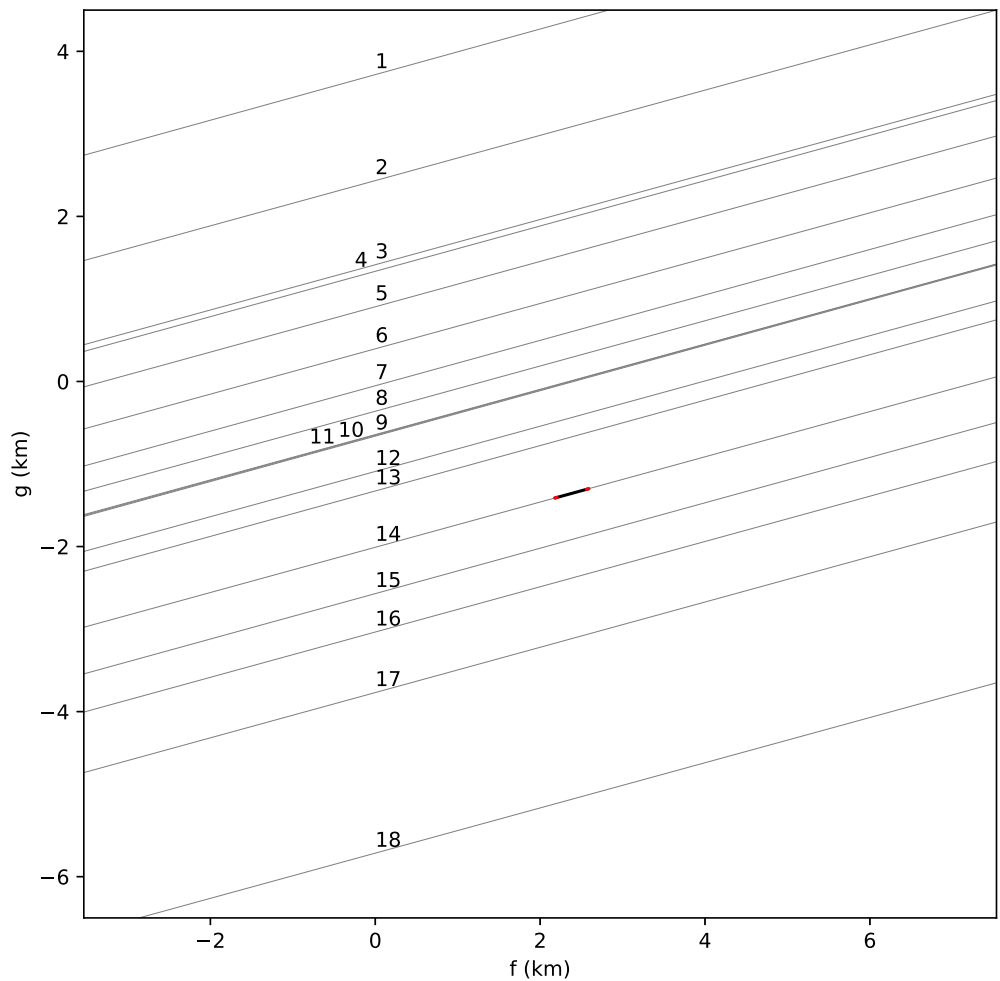

**Fig. 3.** Occultation chords projected on to the sky plane as seen from 18 observation sites. The $f$ and $g$ axes indicate the east–west and north–south directions in the plane, respectively. The site numbers (see table 1) for each chord are also shown. The bold black line represents the positive occultation chord observed at Site No. 14 (Observer: MI). The red segments are the uncertainties (1$\sigma$ level) at the extremities of the positive chord. This chord map is built using the JPL#228 ephemeris (the latest ephemeris for (98943) 2001 CC21 as of 2024 March).

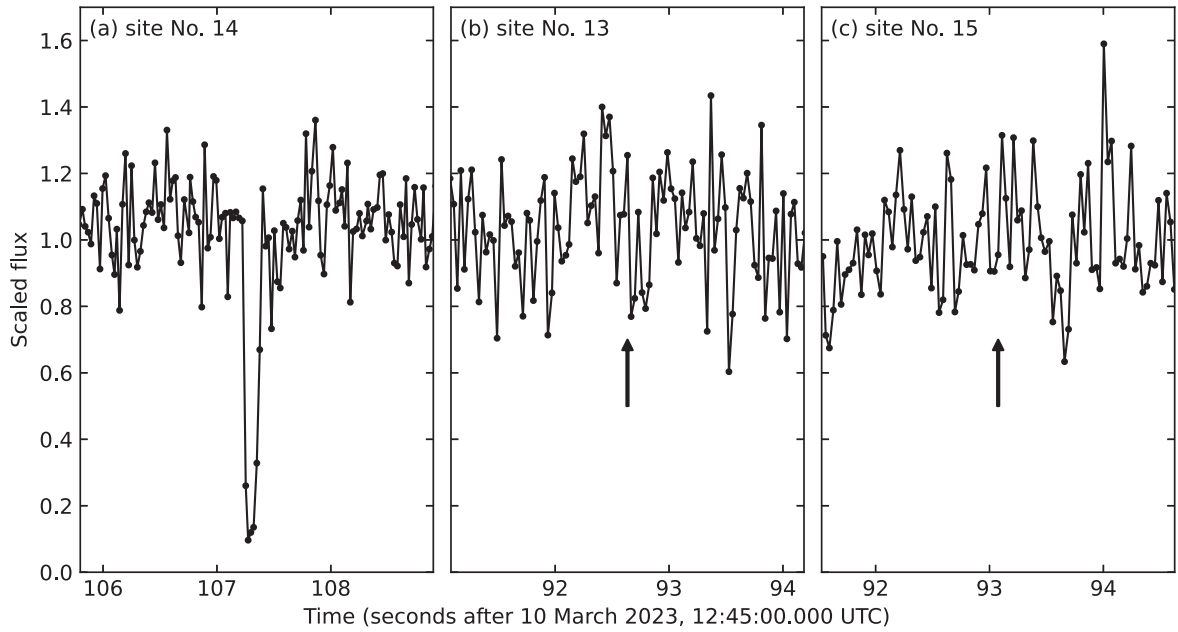

**Fig. 4.** Examples of the light curves. The light curves obtained at (a) Site No. 14 (positive chord, observer: MI) and (b) and (c) its neighbouring chords, Site Nos. 13 (observer: KO) and 15 (observer: TH, see figures 2 and 3). The vertical arrow in panels (b) and (c) shows the time of closest approach to the approximate occultation center estimated from the Site No. 14 occultation data.

The light curve of the occulted star was measured using aperture photometry. The aperture radius was set to ∼1.5–2 times the typical apparent size of the star in the images obtained at each observing site, and the inner and outer annulus radii for the sky background measurements were set to ∼2 and ∼6 times the aperture radius, respectively. The light curves obtained were scaled to the average stellar flux. In case an occultation was detected, they were scaled to the flux observed before and after the event. Figure 4 shows examples of the light curves of TYC 4082-00763-1 obtained in this campaign. Of the 18 reported sites, only one site (Site No. 14 observed by MI, figure 4a) has a positive detection of the occultation. According to the fit of the Fresnel diffraction fringe of an infinite, straight edge of a shadow convolved by a stellar size (radius of 0.07 mas, see subsection 4.2) to the light curve, the apparent immersion and emersion times of the occultation event are $12^\mathrm{h}46^\mathrm{m}47{^\mathrm{s}_\cdot}252 \pm 0{^\mathrm{s}_\cdot}003$ UT and $12^\mathrm{h}46^\mathrm{m}47{^\mathrm{s}_\cdot}346 \pm 0{^\mathrm{s}_\cdot}003$ UT, respectively. These immersion and emersion times correspond to a positive occultation chord with a length of ∼0.40 km. This positive occultation chord projected on to the sky plane is shown in figure 3.

## 4 Results

We have carried out a more detailed analysis using the DOUSHITE procedure on the light curve obtained from the observation of Site No. 14 (figure 4a). Site No. 14 images were obtained using a monochrome CMOS sensor (SONY IMX290), which provides suitable data for accurate photometry. Furthermore, the captured data were not subjected to any output modifications such as gamma value adjustments or sharpening, thus ensuring the integrity of the approximate linearity of the data. The occultation light curve obtained from images with these characteristics is suitable for diffraction analysis with DOUSHITE.

### 4.1 Noise estimation

In order to perform appropriate fitting procedures to the observed light curve of the diffraction features, an accurate characterization of their noise performance is essential. The signal-to-noise ratio ($S/N$) of a star's light curve obtained from high cadence observations using an optical photoelectric detector array can be approximated by the following equation (e.g., Arimatsu et al. 2017):

$$S/N = \frac{I_\mathrm{star}}{\sqrt{\sigma_\mathrm{read}^2 + \sigma_\mathrm{dark}^2 + \sigma_\mathrm{sky}^2 + \sigma_\mathrm{source}^2 + \sigma_\mathrm{sci}^2}}$$
$$= \frac{I_\mathrm{star}}{\sqrt{N_\mathrm{pix}\sigma_\mathrm{back_{pix}}^2 + \sigma_\mathrm{source}^2 + \sigma_\mathrm{sci}^2}}, \qquad (11)$$

where $I_\mathrm{star}$ is the signal value of a star obtained with the aperture area $N_\mathrm{pix}$ pixels ($N_\mathrm{pix} = 28.3$ pixels for the Site No. 14 images) in electrons (hereafter $e^-$). $\sigma_\mathrm{read}$, $\sigma_\mathrm{dark}$, $\sigma_\mathrm{sky}$, $\sigma_\mathrm{source}$, and $\sigma_\mathrm{sci}$ represent the detector readout noise, the dark noise, the sky background noise, the target shot noise, and the scintillation noise, respectively. $\sigma_{\rm read}$, $\sigma_{\rm dark}$, and $\sigma_{\rm sky}$ are expressed as follows:

$$\sigma_{\rm read}^2 = N_{\rm pix} \sigma_{\rm read_{pix}}^2, \quad (12)$$

$$\sigma_{\rm dark}^2 = N_{\rm pix} i_{\rm dark}, \quad (13)$$

$$\sigma_{\rm sky}^2 = N_{\rm pix} s_{\rm sky}, \quad (14)$$

where $\sigma_{\rm read_{pix}}$ is the standard deviation of the detector readout noise in $e^-$ pixel$^{-1}$, $i_{\rm dark}$ is the dark current for the observed exposure time in $e^-$ pixel$^{-1}$, and $s_{\rm sky}$ is the surface brightness of the sky background in $e^-$ pixel$^{-1}$. In equation (11), we introduce the background pixel noise $\sigma_{\rm back_{pix}}$ defined as $\sigma_{\rm back_{pix}}^2 = \sigma_{\rm read_{pix}}^2 + i_{\rm dark} + s_{\rm sky}$. This $\sigma_{\rm back_{pix}}$ value can be approximated by the standard deviation of the background pixel values. $\sigma_{\rm back_{pix}}$ for the Site No. 14 images is estimated to be $\sigma_{\rm back_{pix}} \sim 2.5\ e^-$ pixel$^{-1}$.

---

[3] ⟨https://astro-limovie.info/limovie/limovie_en.html⟩.

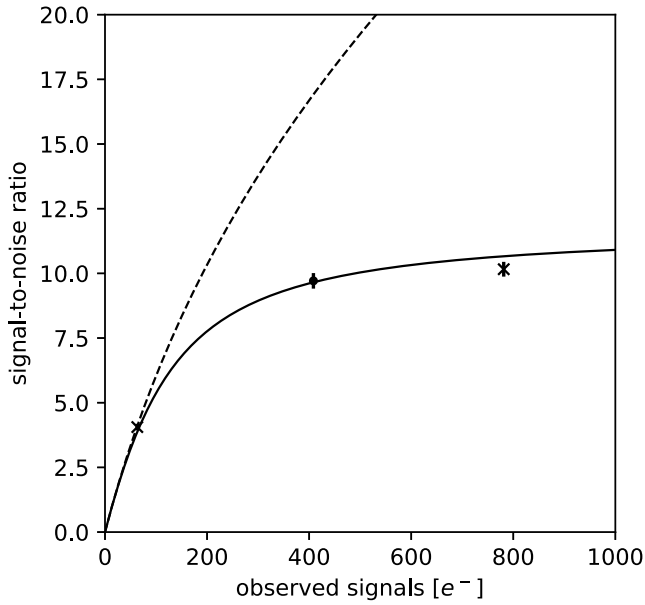

**Fig. 5.** *S/N*s of the light curves for three stars in the Site No. 14 images: the occulted star observed before and after the occultation event (point), and the other two stars (crosses), as a function of their signal values overlaid with the best-fitting synthetic *S/N* values ($\eta = 0.091$, solid line). The dashed line represents $\eta = 0$ for comparison.

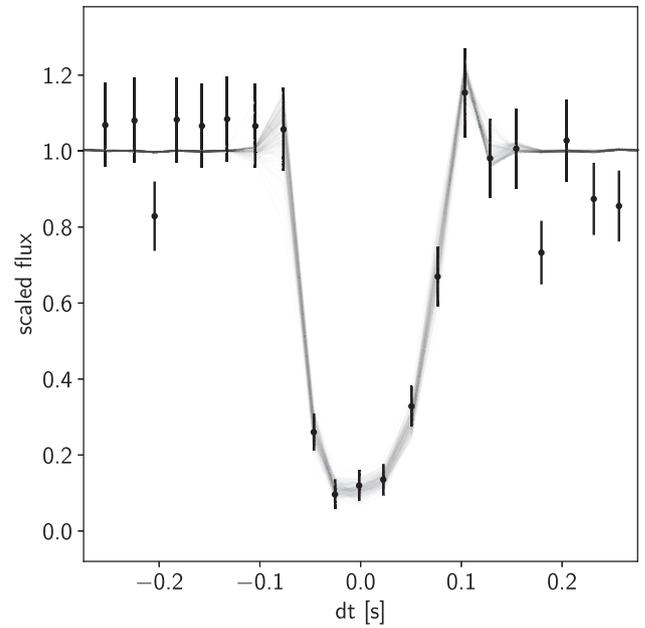

**Fig. 6.** Light curve with error bars obtained from the observation at Site No. 14 (observer: MI). The sample curves obtained from the posterior distributions of the MCMC light curve fit (see figure 8) are superimposed as thin solid lines. The origin of the time ($dt = 0$) is set to the approximate central time of the occultation (2023 March 05 $12^h46^m47\overset{s}{.}300$ UT).

$\sigma_{\rm source}$ and $\sigma_{\rm sci}$ are given by

$$\sigma_{\rm source}^2 = I_{\rm star}, \quad (15)$$

$$\sigma_{\rm sci}^2 = \eta^2 I_{\rm star}^2, \quad (16)$$

where $\eta$ is the fractional scintillation variance relative to $I_{\rm star}$. The synthetic *S/N*s were fitted to those of the light curves for the three stars observed simultaneously in the images: the occulted star observed before and after the occultation event, and the other two stars. The *S/N* of each light curve was estimated as the average value of the stellar fluxes divided by their standard deviation. The *S/N*s of the light curves and the fit results are shown in figure 5. When we assume $\eta = 0$ (dashed line in figure 5), the synthetic *S/N* value is comparable to the observed value for the faintest star, but overestimates it for the brighter ones. This trend indicates that the background noise dominates the main noise component for fainter objects (or the star during the occultation), while the scintillation noise becomes a dominant component for brighter objects. Synthetic *S/N* can be approximated with the observed ones with fractional scintillation variance $\eta = 0.091 \pm 0.009$ (solid line in figure 5). We used the synthetic *S/N* values as the error values for the observed light curve in the following fitting procedures.

Figure 6 shows the light curve of the occulted star with the error bars obtained. The light curve exhibits a significant asymmetry feature, a steep decrease in flux followed by a more prolonged rise, which indicates diffraction effects from a non-spherical occultation shadow.

### 4.2 MCMC fit

In the following, we estimate the shape of the occulting shadow of (98943) 2001 CC21 by considering the effects of diffraction on the obtained light curve. The estimation of the shape is done by determining the posterior probability distribution of the size and shape parameters. We assume that the shadow shape can be approximated by an ellipse with major and minor radii of *a* and *b*, respectively. Figure 7 shows the notation for the size and shape parameters in the sky plane. We define the *x*–*y* plane by aligning the *x*-axis with the occultation chord of Site No. 14 and set the parameters as the centre position of the ellipse, ($x_{\rm c}$, $y_{\rm c}$), its major and minor axes, *a* and *b*, and the apparent position angle. The origin of the *x*–*y* plane coordinates is set to the chord position corresponding to the approximate emersion time (2023 March 5 $12^h46^m47\overset{s}{.}350$ UT). Since shadows with shapes that are axially symmetric to the occultation chord produce identical light curves, our model represents such axially symmetric shapes as equivalent shapes for modelling purposes. There-fore, instead of $y_c$ and the apparent position angle, we introduce the absolute value of $y_c$, $y'_c = |y_c|$, and the corresponding position angle in the $x$–$y'$ plane, PA$'$, as shape parameters (see figure 7).

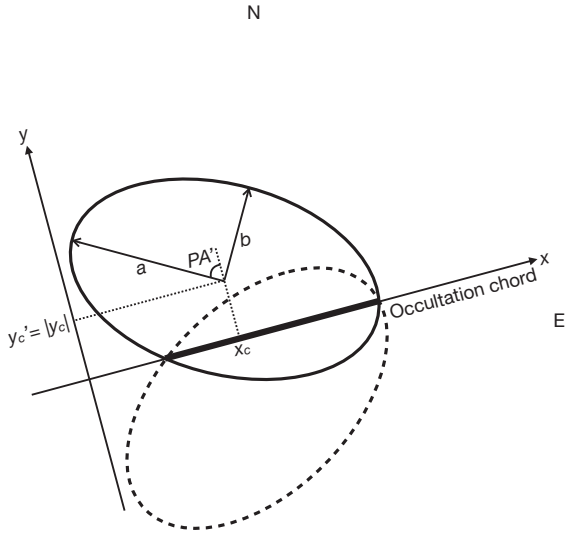

**Fig. 7.** Notation of the model shadow shape used in the present MCMC light curve fit. The bold black line represents the positive occultation chord observed at the Site No. 14 (observer: MI, see also figure 3). The $x$–$y$ plane is aligned with the occultation chord, and the ellipse represents the assumed shape of the occultation shadow with major and minor radii $a$ and $b$, respectively. The central position ($x_c$, $y'_c = |y_c|$) and the position angle (PA$'$) of the ellipse shadow are also key parameters for fitting the model to the observed light curve. See subsection 4.2.

To produce the synthetic light curve $L_i$ (see section 2), the spectral energy distribution [$s(\lambda)$ in equation (8)] for the occulted star TYC 4082-00763-1 is approximated to be a stellar spectral model proposed by Castelli and Kurucz (2003) with $T_{\rm eff} = 5000$ K. This $T_{\rm eff}$ for the selected model is consistent with the estimated effective temperature of TYC 4082-00763-1 from the Gaia DR3 catalog ($T_{\rm eff} = 5134^{+5}_{-4}$ K; Gaia Collaboration 2023). We also adopt the angular radius of TYC 4082-00763-1 to be 0.07 mas, following the Gaia DR3 catalog value. For the spectral efficiency of the observing instrument $e(\lambda)$ we use the quantum efficiency of the SONY IMX290 CMOS sensor available from the camera manufacturer.[4]

To estimate the posterior probability distribution of the size and shape parameters in the Bayesian approach, we employ a Markov chain Monte Carlo (MCMC) scheme. We perform fits of $L_i$ to the observed light curve using the open source ensemble sampler code emcee (Foreman-Mackey et al. 2013). $L_i$ is fitted to the data points of the observed light curve by minimizing $\chi^2$;

$$\chi^2 = \sum_i \frac{(I_i - L_i)^2}{\sigma_i^2}, \qquad (17)$$

where $I_i$ and $L_i$ are the observed and the synthetic stellar flux at the data point $i$, respectively. $\sigma_i$ is the 1$\sigma$ error at $i$ estimated in subsection 4.1. We run 50 chains of MCMC walkers with randomized initial parameters. For each chain we run 12000 steps with a burn-in of the first 2000 steps. The fitting procedure assumes uniform prior distributions for $x_c$ ($-0.5 < x_c < 0.5$ km), $y'_c$ ($0 \leq y'_c < 0.5$ km), $b/a$ ($0 < b/a \leq 1$), and PA$'$ ($0 \leq$ PA$' < \pi$). For the mean radius, $\sqrt{(ab)}$, we adopt a normal prior distribution, anchored to the recent estimates from optical polarimetry and near-infrared spectroscopy observations by Geem et al. (2023). We assume $\sqrt{(ab)} = 0.25$ km reflecting the estimated mean radius, with a standard deviation $\sigma_{\sqrt{(ab)}} = 0.05$ km, to account for the observational uncertainties. When fitting the model, we found that several chains became entrenched in parameter regions of low probability. To exclude such chains from subsequent analyses, we used a selection method based on the approach proposed by Nir et al. (2023) with the following criterion

$$\langle \chi^2 \rangle - {\rm median}(\langle \chi^2 \rangle) > 2.5\,{\rm MAD}_{\chi^2}, \qquad (18)$$

where $\langle \chi^2 \rangle$ is the average mean of the $\chi^2$ values for all points for each chain, and ${\rm MAD}_{\chi^2}$ is the median absolute deviation (MAD) of $\chi^2$, respectively. Chains with $\langle \chi^2 \rangle$ that met this criterion were excluded from the following analyses.

---
[4] ⟨https://astronomy-imaging-camera.com/wp-content/uploads/ASI290MM-QE.jpg⟩.

Figure 8 shows the posterior distributions of the parameters for the 2023 March 5 event obtained from the MCMC fit, and the resulting best-fitting parameters with their uncertainties are listed in table 2. We should note that the probability $b/a$ has a bimodal distribution. More than 88% of the probability is located in one mode around a peak at 0.36, and the posterior has a second mode at 0.18 (approximately half the size of the first peak) that contains $\sim 12$% of the probability. However, the second mode does not significantly affect the resulting best-fitting value of $b/a = 0.37 \pm 0.09$. All other parameters are well constrained to a single mode. The best-fitting PA$'$ value (PA$' = 108^{\circ+4}_{-5}$) in the $x$–$y'$ coordinates corresponds to two possible position angles of the major axis in the celestial coordinates, $124^{\circ+4}_{-5}$ and $87^{\circ+5}_{-4}$. The best-fitting $(x, y') = (-0.07^{+0.06}_{-0.05}, 0.14 \pm 0.04)$ km is consistent with two possible solutions for the central coordinate of (98943) 2001 CC21 in the projected $f$–$g$ sky plane relative to its predicted position with the latest JPL #228 ephemeris as of 2024 March (see figure 3), $(f, g) = (2.49^{+0.05}_{-0.04}, -1.18 \pm 0.05)$ km and $(f, g) = (2.56^{+0.05}_{-0.04}, -1.45 \pm 0.05)$ km. However, as already noted in section 3, the astrometry of the occulted star is possibly inaccurate, and these coordinate values should be used with caution when revising the ephemeris of the asteroid.

Figure 6 shows the sample light curves obtained from the posterior distributions. The sample curves reproduce the asymmetry feature in the observed light curve well. The sample elliptical shadow shapes of (98943) 2001 CC21 obtained from the posterior distributions are shown in figure 9. Two modes of the posterior shapes corresponding to the bimodal $b/a$ distribution are clearly visible. The maximum $y'$ values of approximately 99.8% of the posterior shapes are less than 0.6 km and are consistent with the non-detection of flux variations at the neighboring chords, Sites No. 13 and No. 15 (figures 4b and 4c), whose distances from Site No. 14's chord are 0.61 and 0.53 km, respectively (see also figure 3).

## 5 Discussions

The most probable semi-axes ratio of the elliptical occultation shadow estimated in this study ($b/a \sim 0.37$) indicates a possible elongated shape for (98943) 2001 CC21, which is consistent with its observed light curve with a large amplitude of $\sim 0.82$ mag (Warner 2023). The possible axial ratio implies that the eccentricity of (98943) 2001 CC21 is comparable to or greater than that of (25143) Itokawa ($b/a = 0.55$ and $c/a = 0.39$; Fujiwara et al. 2006). The present study provides the constraint on the projected occultation shadow of (98943) 2001 CC21. However, we detected it only in a single occultation event. Further occultation observations are needed to obtain its exact three-dimensional shape.

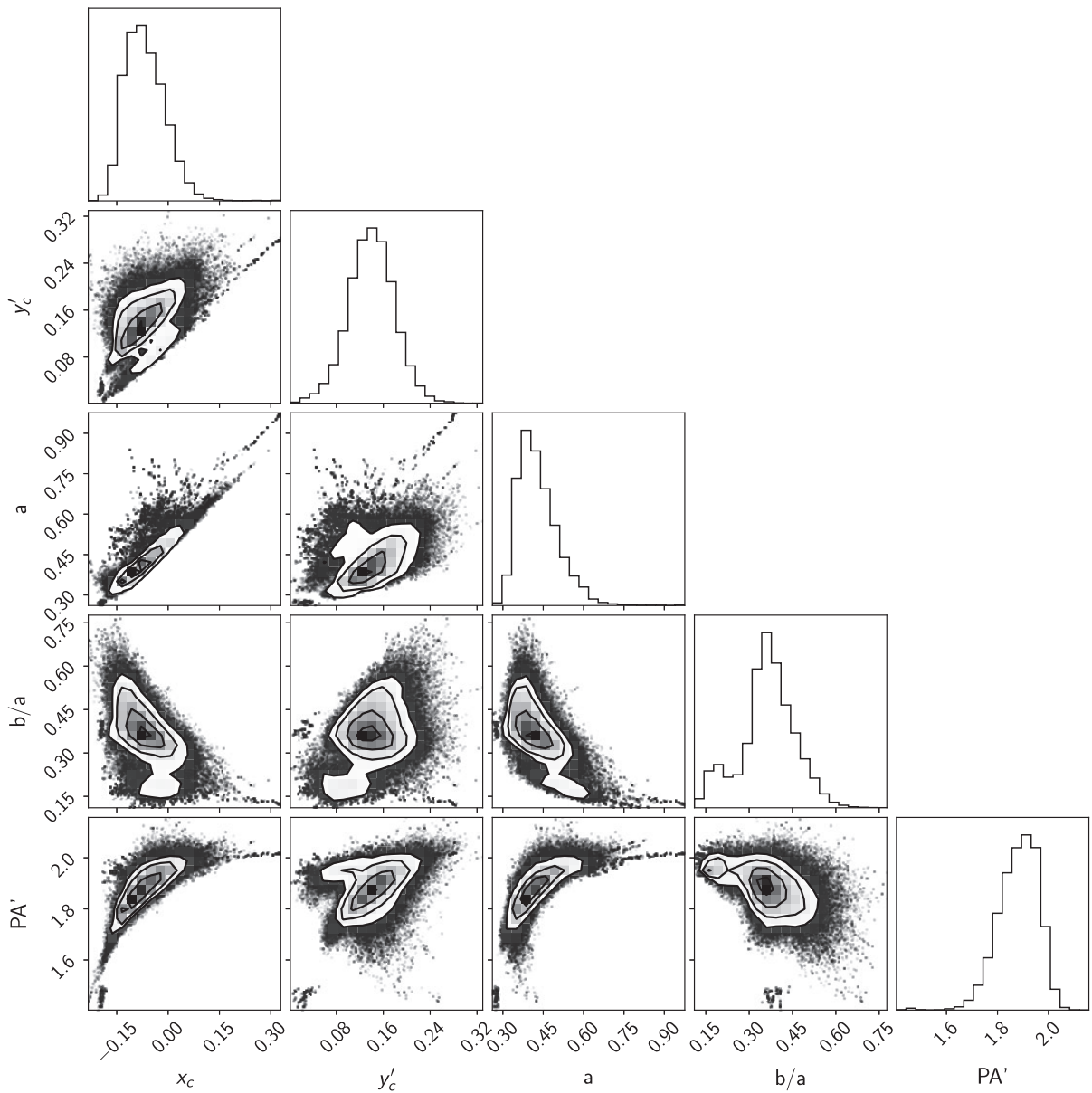

**Fig. 8.** Posterior distributions of the parameters derived from the MCMC light curve fit.

**Table 2.** Best-fitting parameters.

| | |
|---|---|
| $x$ [km]    | $-0.07^{+0.06}_{-0.05}$ |
| $y'$ [km]   | $0.14 \pm 0.04$ |
| $a$ [km]    | $0.42^{+0.08}_{-0.06}$ |
| $b/a$       | $0.37 \pm 0.09$ |
| $PA'$ [°]   | $108^{+4}_{-5}$ |

As mentioned in section 1, imaging opportunities during the Hayabusa2 spacecraft's flyby at (98943) 2001 CC21 will be limited. Given the potential elongation of (98943) 2001 CC21, it is crucial to develop observation strategies that can accommodate a wide range of phase angles. This will involve optimizing camera angles and exposure settings to capture detailed images of the asteroid's surface, regardless of its orientation at the time of the flyby. In turn, understanding the spin-pole orientation is useful for predicting the asteroid's rotation and identifying the observable area during the flyby. This knowledge will allow us to maximise the scientific return from the imaging opportunities, despite the limitations imposed by the high-speed nature of the encounter. Future occultation observations, as well as optical photometric monitoring, should therefore focus on estimating these rotation characteristics.

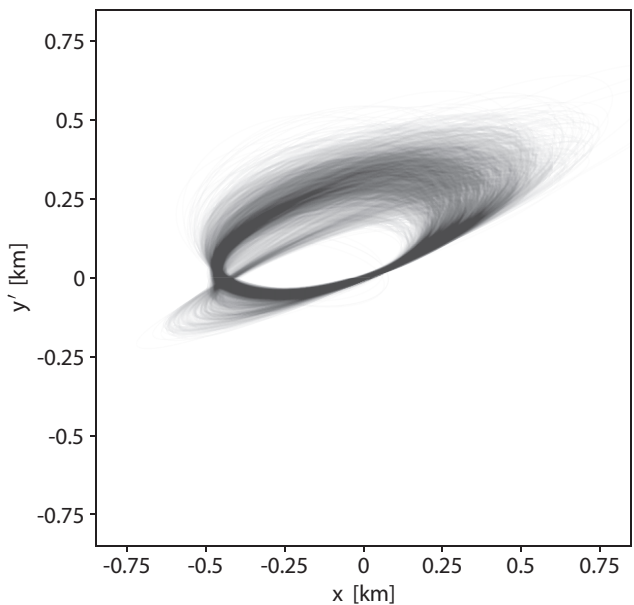

**Fig. 9.** Sample elliptical shapes of (98943) 2001 CC21's occultation shadow obtained from the posterior distributions of the MCMC light curve fit.

## 6 Summary and prospects

Using synthetic light curves produced with the DOUSHITE occultation data reduction procedure, we have successfully obtained observational constraints on the shape of the shadow of (98943) 2001 CC21 from a single chord of occultation observations. The best-fitting parameters indicate that the (98943) 2001 CC21 shadow has an elongated shape with an axis ratio of $b/a = 0.37 \pm 0.09$. During the Hayabusa2 spacecraft's high-speed flyby of (98943) 2001 CC21, observation strategies must be carefully tailored to its potentially elongated shape and varying phase angles, given the constraints imposed by the asteroid's spin-pole orientation and limited imaging capabilities.

Despite the significant improvements in astrometry provided by the Gaia catalog, the observation of stellar occultation events caused by subkilometer- to kilometre-sized bodies remains extremely challenging. In fact, as of 2024 mid-January, 12 observing campaigns of stellar occultations by (98943) 2001 CC 21 have been conducted. However, the only successful detection of a change in stellar brightness due to these occultations is the single case reported in this paper. Future observations of NEAs and other small solar system bodies will be conducted to understand their physical properties and to make flyby missions feasible. However, it is expected that many of these will result in successful observations of at most single or a few chords. The analytical method developed in this study, in which observed light curves are compared with models taking into account diffraction effects, is proving useful in estimating the size and shape of small bodies from such sparse observational data.

## Acknowledgments

We extend our gratitude to the anonymous referee for their meticulous review and the provision of constructive suggestions. This research has been partially supported by JSPS grants (18K13606, 21H01153).